\input{aipcheck}

\documentclass[
   ,final            
  ]
  {aipproc}

\layoutstyle{6x9}

\begin{document}

\title{Looking for black-holes in X-ray binaries with XMM-Newton: XTE~J1817-330 and XTE~J1856+053}



\author{Gloria Sala}{
  address={Max-Planck-Institut f\"ur extraterrestrische Physik, 
  PO Box1312, 85741 Garching b.M., Germany}
}

\author{Jochen Greiner}{
  address={Max-Planck-Institut f\"ur extraterrestrische Physik, 
  PO Box1312, 85741 Garching b.M., Germany}
}

\author{Natalia  Primak}{
  address={Max-Planck-Institut f\"ur extraterrestrische Physik, 
  PO Box1312, 85741 Garching b.M., Germany}
}

\begin{abstract}
The X-ray binary XTE~J1817-330 was discovered in outburst on 26 January 2006 with
RXTE/ASM. One year later, another X-ray transient discovered in 1996, XTE~J1856+053, was
detected by RXTE during a new outburst on 28 February 2007. We triggered XMM-Newton 
target of opportunity observations on these two objects to constrain their 
parameters and search for a stellar black holes. 
We summarize the properties of these two X-ray transients and show that the soft
X-ray spectra indicate indeed the presence of an accreting stellar black hole in each of the two 
systems.
\end{abstract}

\maketitle


\section{Introduction}

X-ray binaries are the brightest X-ray sources in the sky. They are powered
by the accretion of material from the secondary star 
onto a compact object (neutron star or black hole). 
The accretion regime depends on the spectral type of the companion. 
In high-mass X-ray binaries (HMXB)
the secondary star is an O or B star with a strong stellar wind which is 
intercepted and accreted by the compact object. 
Though an accretion disk may also be present,
the secondary star is dominating the optical emission of the source.
In low-mass X-ray binaries (LMXB), the secondary star is of a 
type later than A and the accretion occurs through Loche robe overflow.  
The optical emission during outburst is in this case dominated by the X-ray heated companion,
the outer disk and/or reprocessed hard X-rays. 

The generally accepted picture for the X-ray emission of accreting black holes consists of 
an accretion disk, responsible for thermal black body emission in the X-ray band;
and a surrounding hot corona, origin site of non-thermal power-law emission, 
up to the energy range of gamma-ray telescopes, due to inverse comptonization 
of soft X-ray photons from the accretion disk.

At present, around 20 X-ray binaries contain a dynamically confirmed black hole, and 
around another 20 are the so called black-hole candidates \cite{rmc06}. 
Seven of the 20 confirmed black holes, and 12 of the black-hole candidates
are transient sources with only one unique outburst observed.

\section{The X-ray transients XTE~J1817-330 and XTE~J1856+053}

XTE~J1817-330 was discovered by the Rossi X-ray Timing Experiment (RXTE) on
26 January 2006 \cite{rem06} with a flux of 0.93($\pm0.03$)~Crab (2-12 keV) and 
a very soft spectrum, typical for black hole transients.
The RXTE light-curve of XTE~J1817-330 reached a maximum of 1.9~Crab
on 2006 January 28, and then declined exponentially, with an e-folding time of 27 days.
Following the initial discovery, the radio, near infrared and optical counterparts were identified  
\cite{rup06a,rup06b,dav06,tor06}.

The X-ray transient XTE~J1856+053 was discovered with RXTE/PCA during a survey of the Galactic ridge 
in 1996 \cite{mar96}. The RXTE/ASM light-curve of the 1996 outburst showed two peaks (Fig.~\ref{lcs}):
a first symmetric one starting on 1996 April 4, 27 days long in total and with a maximum of  75~mCrab (2--12~keV);
and a second fast rise--slow decay (FRED) peak starting on 1996 Sept. 9, lasting for 70 days, and reaching 
a maximum flux of 79~mCrab \cite{rem99}. The second X-ray peak was preceded 8 days before 
by a high-energy (20--100~keV) precursor of 30--60~mCrab detected by BATSE on 1996 Sept. 7--9 \cite{bar96}. 
A new outburst of XTE~J1856+053 was detected on 2007 February 28 \cite{lev07}.
As in 1996, the RXTE/ASM light curve of the 2007 outburst shows two peaks, 
preceded by a precursor on 2007 January 10--15 (Fig.~\ref{lcs}). The first peak reached 
a maximum of $\sim$85~mCrab on  March 12 and lasted for $\sim$65~days. The second peak 
in 2007 rose in only 7 days to $\sim$110~mCrab, and lasted for $\sim$55~days.
As in the case of the second maximum in 1996, the two 2007 peaks were preceded by hard X-ray precursors 
detected by Swift/BAT (10--200~keV) in the periods February 22 to March 1, and 2007 May 28--30 \cite{kri07}.

\section{XMM-Newton observations}

Short after the start of the outburst of XTE~J1817-330, 
we triggered a Target of Opportunity Observation (TOO) 
with XMM-Newton (0.1-10.0~keV) which was executed on the first XMM-Newton 
visibility window of the source, on 2006 March 13 (obs. ID. 0311590501, 20~ks), 
when the source flux detected by the All Sky Monitor (ASM) on board RXTE had faded to $\sim$300~mCrab \cite{sal06,salsub}.

\begin{figure}
\label{lcs}
  \includegraphics[height=.24\textheight]{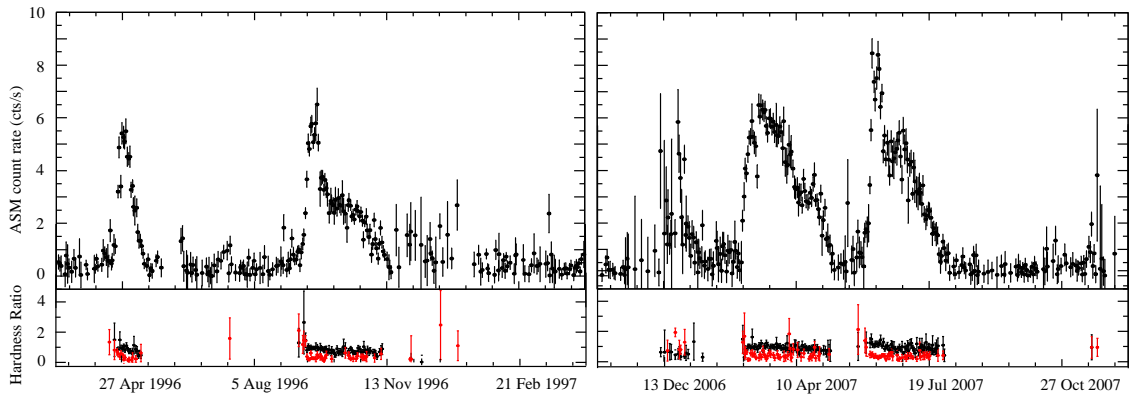}
  \caption{RXTE/ASM light-curves and hardness ratios of XTE J1856+053 during its 1996 and 2007 outbursts.
The upper panels show the light-curves in the total band (1.5--12~keV). 
The lower panels show the two hardness ratios: (3--5~keV)/(1.5--3~keV) 
is shown in black and (5--12~keV)/(3--5~keV) is shown in red.}
\end{figure}

In a similar way, we obtained a TOO observation of XTE J1856+053 with XMM-Newton (0.1-10.0~keV), 
on 14 March 2007 (obs. ID. 0510010101, 5~ks for RGS, 1.5~ks for EPIC-pn), 
when the source flux detected by the RXTE/ASM was in its first maximum, $\sim$80~mCrab \cite{sal07,sal08}(Fig.~\ref{lcs}).

Due to the high brightness of the source, the EPIC-pn camera on board XMM-Newton 
(0.2--10.0~keV) was used in the burst mode for XTE~J1817-330 and in timing mode for XTE~J1856+053 
to avoid pile-up, while the Reflection Grating Spectrometer (RGS,0.3--2.0~keV) 
was operated in spectroscopy high count rate mode. 
The Optical Monitor (OM) was used in fast mode with the U and UWV1 filters for XTE~J1817-330,
and with the U filter for XTE~J1856+053.

XMM-Newton data were reduced with the XMM-Newton Science Analysis Software v7.1,
and XSPEC 11.3 was used for spectral analysis. In the case of XTE~J1817-330, the source 
is clearly detected with the OM both in U and UVW1, with AB magnitudes U$=15.952\pm0.01$ and UVW1$=16.13\pm0.01$. 
The EPIC-pn and RGS spectra are fit together in this case with the OM U and UWV1 data, 
which provide an extra constraint to the absorption and the accretion disk. 
In the case of XTE~J1856+053, the source is not detected in the OM images, with an upper limit for the AB magnitude
in the U band during outburst of 23 mag.

\section{Results}

We use a 2-component model consisting of a thermal accretion disk 
plus a comptonization component ({\it compTT} model \cite{tit94}) 
to fit the spectra of XTE~J1817-330 and XTE~J1856+053.
The standard thermal accretion disk model is composed by the sum of black bodies 
from surface temperatures depending on the radius as $R^{-3/4}$ 
\cite{mit84} (\emph{diskbb} model in \emph{xspec}). 
Since this model neglects the torque-free boundary condition, 
the temperature distribution is not accurate for the innermost disk when it extends down to the 
innermost stable orbit. The \emph{diskpn} model in \emph{xspec} \cite{gie99}
includes corrections for the temperature distribution near the black hole 
by taking into account the torque-free inner-boundary condition.
We use here the {\it diskpn} model, fixing the inner radius to 6$R_g$.

For XTE~J1817-330, we obtain the best fit for the {\it diskpn + compTT} model ($\chi^2_{\nu}=1.18$) with 
$N_{\rm{H}}=1.55(\pm0.05)\times10^{21}\,\mbox{cm}^{-2}$,
a disk with $kT_{\rm{in}}=0.70(\pm0.01)\,\mbox{keV}$ and a
comptonization component with kT$_{e}=50$~keV (fixed) and $\tau=0.15(\pm0.02)$.
The observed X-ray flux is 
$8.6(\pm 0.8)\times10^{-9}\,\mbox{erg}\,\mbox{cm}^{-2}\,\mbox{s}^{-1}$ (0.4-10~keV), 
and the unabsorbed X-ray luminosity of the source at the time of the observation 
L$_{(0.4-10\,keV)}=1.2(\pm0.1)\times10^{38}({D}/10\rm{kpc})^2\,\mbox{erg}\,\mbox{s}^{-1}$.

In the case of XTE~J1856+053, no indication of a hard component is evident in the residuals, 
but an excess is present below 1 keV, leading to a poor reduced 
$\chi^2$ of 1.99. Adding a recombination emission edge
at 0.87~keV (corresponding to O~VIII~K-shell) with
plasma temperature kT$=50(\pm3)$~eV improves the fit.
The significance of this feature is however to be taken with 
care, since the excess could be caused by some redistribution of 
higher energy photons to lower energies not properly 
taken into account by the calibration.
The best fit ($\chi^2_{\nu}=1.16$) is then obtained with 
$N_{\rm{H}}=4.5(\pm0.1)\times10^{22}\,\mbox{cm}^{-2}$, and
a disk ({\it diskpn} model) with $kT_{\rm{in}}=0.75(\pm0.01)\,\mbox{keV}$.
The observed X-ray flux is 
$1.0(\pm0.1)\times10^{-9}\,\mbox{erg}\,\mbox{cm}^{-2}\,\mbox{s}^{-1}$ 
(0.5--10.0~keV), which corrected for absorption corresponds to an unabsorbed X-ray luminosity 
L$_{(0.5-10.0\,keV)}=4.0(\pm1.5)\times10^{38}({D}/10\rm{kpc})^2\,\mbox{erg}\,\mbox{s}^{-1}$.

\section{Discussion}

The low temperature of the accretion disk favours a black-hole as the 
accreting compact object both in XTE~J1817-330 and XTE~J1856+053.
An upper limit for the compact object mass can be obtained from the X-ray spectrum.
The normalization constant {\it K} of the {\it diskpn} model is 
related to the mass of the compact object {\it M}, the 
distance to the source {\it D}, and the inclination {\it i} of the disk as 
$K=\frac{M^2 cos(i)}{D^2 \beta^4}$,
where $\beta$ is the color/effective temperature ratio. 
Furthermore, the accretion rate can be obtained
from the mass of the compact object and the maximum temperature of the disk \cite{gie99}. 

Assuming $\beta=1.7$ and using the best fit value for the normalization of the 
{\it diskpn} model for XTE~J1817-330 ($K_{\rm {diskpn}}=0.024\pm0.002$), we can compare the accretion rate 
for different possible masses, distances, and inclinations with an
upper limit for the accretion rate. 
At the time of the XMM-Newton observation of XTE~J1817-330, the flux of the source 
had decreased by a factor 6 with respect to the maximum registered by RXTE. 
Taking the Eddington limit as the upper limit for 
the accretion rate at the maximum of the outburst,
the accretion rate at the time of the observation could not be higher than 
16\% of $\rm M^{\rm{acc}}_{\rm{Edd}}$.
This sets an upper limit for the mass of the central object of 6~M$_{\odot}$
(see \cite{salsub} for more details).

In the case of XTE~J1856+053, $K_{\rm {diskpn}}=(8.5\pm0.4)\times10^{-3}$. 
At the time of the XMM-Newton observations, the flux of the source 
was at the maximum of the first outburst detected in March 2007. However, a brighter
outburst was detected by RXTE/ASM in June 2007. At the time of our XMM-Newton observations, the flux was 70\% 
of the maximum detected in June 2007. 
Taking the Eddington limit as the upper limit for the accretion rate at the 
maximum in June 2007, the accretion rate at the time of the XMM-Newton 
observation could not be higher than  $0.7\rm M^{\rm{acc}}_{\rm{Edd}}$.
This sets an upper limit for the mass of the central object of 4.2~M$_{\odot}$. 

We note, however, that our estimation for the upper limits of the compact object mass 
depends on the color correction factor.
The value adopted here ($\beta=1.7$) is rather conservative in comparison, for example, 
to $\beta=2.6$ obtained by Shrader \& Titarchuk \cite{shra03}. 
A larger color correction factor would, of course, 
imply a larger upper limit for the black hole mass.

\begin{figure}
\label{mm}
  \includegraphics[height=.24\textheight]{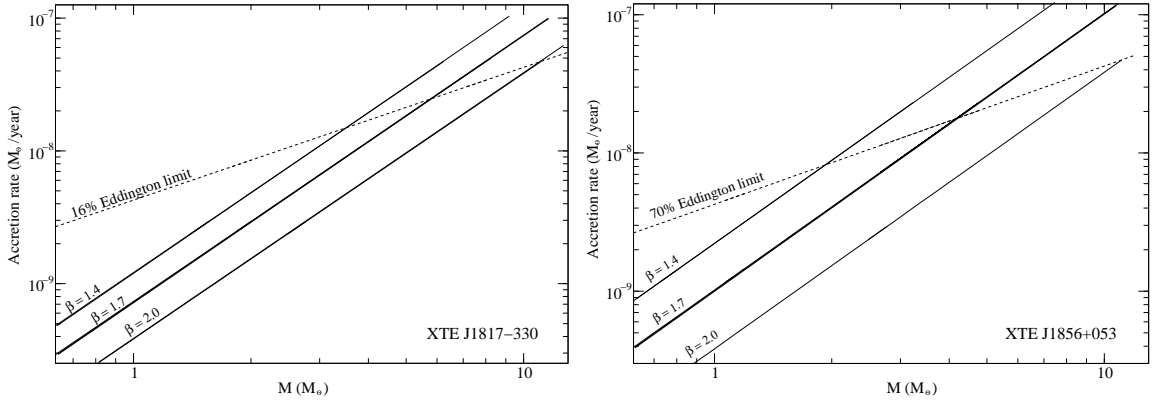}
  \caption{Accretion rate as a function of the black hole mass derived from the X-ray spectral fit of
XTE~J1817-330 (left) and XTE~J1856+053 (right). Over plotted is shown the accretion rate upper limit
obtained by assuming that at maximum the source was at the Eddington limit and scaling it down 
to the X-ray flux at the time of the XMM-Newton observation.}
\end{figure}

The absorption column obtained for XTE~J1817-330
is in agreement with the low absorption already pointed out at the 
discovery of the source with RXTE \cite{rem06}), 
and confirmed later in UV and X-rays \cite{mil06a,ste06,mil06b,gol06}.
The low absorption towards XTE~J1817-330 makes this new black hole candidate
an ideal target for optical observations in the quiescent state, which could provide the 
dynamical confirmation of the presence of a black hole in the system.

The type of the secondary star can be constrained from the non-detection of 
the sources in quiescence. XTE~J1817-330 was not detected 
previous to the outburst in the digitized sky survey (DSS, V$>$22~mag).  
With the extinction towards the source derived from the N$_{\rm H}$, A$_{\rm V}=0.76$
(using $N_{\rm H}=5.9\times10^{21}E_{B-V}\,\mbox{cm}^{-2}$ \cite{zom07}), 
the lower limit on the absolute magnitude is M$_{\rm V}> 6$~mag (for 10~kpc). 
A limit on the absolute magnitude of M$_{\rm V}> 6$~mag implies 
that the secondary star must be a K-M star, and a giant can be excluded, 
even for a 10~kpc distance. 
We therefore conclude that XTE~J1817-330 is a low-mass X-ray binary.

XTE~J1856+053 was observed on 26 August 2007 with GROND \cite{grond}, 
a 7-channel imager mounted at the MPI/ESO 2.2m telescope at La Silla (Chile).
The source was not detected in a 8-minute exposure, with lower limits for the IR magnitudes
of J$\sim$20, H$\sim$19, and K$\sim$18 (calibrated against 2MASS). 
This rules out the presence of a massive secondary in XTE~J1856+053. 
Even for a distance of 10~kpc, an O or B star would be detected with J$\sim$9--11. 
We can thus classify XTE~J1856+053 as a low mass X-ray binary (LMXB). 
We can also use the GROND non-detection to estimate a lower limit for the distance:
for a faint M2 secondary (the latest spectral type of a donor in a confirmed BH binary;
see Tab. 1 in McClintock \& Remillard \cite{mcrem04}), and with the extinction derived from
the X-ray absorption column (A$_J$=6.7, using $N_{\rm H}=5.9\times10^{21}E_{B-V}\,\mbox{cm}^{-2}$
and  A$_J$=0.282A$_V$ \cite{zom07,rl85})
the non-detection in the J band would set a lower limit for the distance of 1~kpc.


\begin{theacknowledgments}
We thank Norbert Schartel and the XMM-Newton team for carrying out the 
TOO observations presented here. We acknowledge
the RXTE/ASM team for the public availability of the quick-look results.
XMM-Newton is an ESA Science Mission
directly funded by ESA Member States and the USA (NASA),
with support from BMWI/DLR (FKZ 50 OX 0001) and the Max-Planck
Society.
GS is supported through DLR (FKZ 50 OR 0405).
\end{theacknowledgments}



\bibliographystyle{aipprocl} 




\end{document}